\documentclass[twocolumn]{aastex62}

\usepackage{amsmath}
\usepackage{amssymb}
\usepackage{graphicx,epstopdf}
\usepackage{hyperref}
\usepackage[normalem]{ulem}

\begin{document}

\title{Constraints on the moment of inertia of PSR J0737-3039A from GW170817}

\shorttitle{Constraints on the moment of inertia of PSR J0737-3039A from GW170817}
\shortauthors{Landry and Kumar}

\email{landryp@uchicago.edu, bharat@iopb.res.in}

\author{Philippe Landry}
\affiliation{Enrico Fermi Institute and Kavli Institute for Cosmological Physics, \linebreak The University of Chicago, 5640 South Ellis Avenue, Chicago, Illinois, 60637, USA}
\author{Bharat Kumar}
\affiliation{Inter-University Centre for Astronomy and Astrophysics, Post Bag 4, 
Ganeshkhind, Pune, 411007, India}
\affiliation{Institute of Physics, HBNI, Sachivalaya Marg, Bhubaneswar, 751005, India}

\received{--}
\revised{--}
\accepted{--}
\submitjournal{ApJL}

\begin{abstract}
Continued observation of PSR J0737-3039, the double pulsar, is expected to yield a precise determination of its primary component's moment of inertia in the next few years. Since the moment of inertia depends sensitively on the neutron star's internal structure, such a measurement will constrain the equation of state of ultra-dense matter, which is believed to be universal. Independent equation-of-state constraints have already been established by the gravitational-wave measurement of neutron-star tidal deformability in GW170817. Here, using well-known universal relations among neutron star observables, we translate the reported 90\%-credible bounds on tidal deformability into a direct constraint, $I_{\star} = 1.15^{+0.38}_{-0.24} \times 10^{45} \text{ g cm}^2$, on the moment of inertia of PSR J0737-3039A. Should a future astrophysical measurement of $I_{\star}$ disagree with this prediction, it could indicate a breakdown in the universality of the neutron-star equation of state.
\end{abstract}


\section{\label{sec:intro}Introduction.} 

PSR J0737-3039 is the only double pulsar known to date \citep{Burgay03,Lyne04}. Thanks to precision timing of the radio pulses from its $1.338\,M_{\odot}$ primary component (PSR J0737-3039A, hereafter pulsar A), many of the system's post-Keplerian parameters, describing relativistic corrections to the orbital motion, are well-measured \citep{Kramer}. In particular, the periastron advance has been determined to better than 1 part in $10^4$. Part of the advance is due to relativistic spin-orbit coupling \citep{Damour,Wex}, and forthcoming improvements in the measurement of the orbital decay will permit the spin correction to be distinguished from the standard post-Newtonian advance \citep{KramerWex}. The measurement of pulsar A's spin angular momentum $S$ is expected to determine its moment of inertia $I_{\star}$ with $\sim 10\%$ accuracy in the next few years \citep{Lyne04,Lattimer05,KramerWex}. A moment of inertia measurement is highly anticipated because of its ability to constrain the neutron star \emph{equation of state}---the pressure-density relation inside the star---in the high-density regime \citep{Morrison04,Lattimer05,Bejger05,Worley,Raithel,Gorda}; such a constraint would have implications for the mass distribution of astrophysical neutron stars, the end state of binary mergers, and r-process nucleosynthesis, among other questions \citep{OzelFreire}.

The macroscopic properties of neutron stars, including the moment of inertia $I$, depend strongly on the characteristics of ultra-dense matter encoded in the equation of state. Since the supranuclear densities attained in the core of a neutron star are beyond the reach of laboratory experiments, the equation of state is poorly constrained above $\rho_{\text{nuc}} \approx 2.8 \times 10^{14} \text{ g cm}^{-3}$. Competing models from nuclear theory disagree on the structure and composition of the core: predicted densities vary by nearly an order of magnitude, and the abundances of exotic particles like hyperons or free quarks are uncertain \citep{OzelFreire}. Astrophysical observations of neutron stars are critically important for resolving these disagreements.

One observational approach that has recently borne fruit consists of gravitational-wave measurement of the neutron-star tidal deformability, $\Lambda$. The tidal deformability is an intrinsic stellar property that determines how easily a star is deformed by tidal forces. It correlates strongly with the \emph{stiffness} of the equation of state, i.e.~the size of the pressure gradients inside the star. In a binary, the gravitational field of a neutron star's companion generically raises a stellar quadrupole moment whose amplitude is proportional to $\Lambda$. The tidal bulge sources gravitational radiation, dissipating energy and slightly accelerating the coalescence relative to the merger of point-particles. The net effect on the waveform is a small but measurable $\Lambda$-dependent phase shift (see e.g.~\citet{Flanagan,Hinderer,ReadGW,DelPozzo,Wade} and references therein).

The amplitude of the tidal phase shift was constrained by Advanced LIGO's \citep{LIGO} and Virgo's \citep{VIRGO} detection of GW170817, a loud compact binary merger signal identified \citep{LIGOsearch1,LIGOsearch2} in gravitational-wave strain data \citep{LIGOcalibration} and determined to be astrophysical in origin \citep{LVCdetection}. After removing a noise transient present in LIGO data \citep{LIGOglitch2}, analysis of the properties of the source \citep{LIGOpe} yielded the first observational bounds on tidal deformability, which were presented with the discovery \citep{LVCdetection}. The initial analysis of the tidal phasing in \citet{LVCdetection} set a 90\%-credible upper bound of $\Lambda_{1.4} \leq 800$ on the tidal deformability of a $1.4\,M_{\odot}$ neutron star. That study did not explicitly assume that both compact objects were neutron stars, as the gravitational-wave data alone could not rule out the possibility of a neutron-star black-hole merger. A subsequent analysis, which assumed---based on electromagnetic data and other indicators---that the binary consisted of neutron stars with the same equation of state \citep{Chatziioannou,Carney}, tightened the bounds to $\Lambda_{1.4} = 190^{+390}_{-120}$ at 90\% confidence \citep{LVCeos}. (A separate study \citep{De} also improved upon the original inference, but did not publish direct constraints on $\Lambda_{1.4}$.) \citet{LVCeos}'s bounds are the most stringent constraints on neutron-star tidal deformability reported to date.

Because the neutron-star equation of state is believed to be universal, the tidal deformability constraints from GW170817 have implications for \emph{all} neutron stars, including PSR J0737-3039A. In this letter, the $90\%$ confidence interval on $\Lambda_{1.4}$ from GW170817 is translated into a direct constraint on the moment of inertia of pulsar A. The conversion relies on the existence of \emph{universal relations} for neutron stars \citep{YagiILQ,YagiILQscience}, functional relationships between pairs of internal-structure dependent observables that turn out to be approximately insensitive to the equation of state (see \citet{YagiILQreview} for a review). Two types of universal relation are employed in this work. The \emph{binary Love relation} \citep{YagiBiLove,YagiBiLove2} between the tidal deformabilities of two neutron stars of different masses is used to map the $\Lambda_{1.4}$ constraints to $\Lambda_{\star}$ bounds, where $\Lambda_{\star}$ is the tidal deformability of a $1.338\,M_{\odot}$ star, like pulsar A. The \emph{I-Love relation} \citep{YagiILQ,YagiILQscience} between the dimensionless moment of inertia $\bar{I}:=c^4I/G^2 M^3$ and the tidal deformability $\Lambda$ is used to convert the $\Lambda_{\star}$ bounds to a 90\% confidence interval on $I_{\star}$. Applying the relations to the gravitational-wave tidal constraints quoted above, GW170817 is found to constrain pulsar A's moment of inertia to \mbox{$I_{\star} = 1.15^{+0.38}_{-0.24} \times 10^{45} \text{ g cm}^2$}. This figure accounts for the error associated with the approximate nature of the universal relations, and explicitly relies on the identification of GW170817 as a binary neutron star merger. The less restrictive upper bound of $\Lambda_{1.4} \leq 800$ from the initial analysis of GW170817 corresponds to $I_{\star} \leq 1.67 \times 10^{45}~\text{g cm}^2$.

\begin{figure}[t]
\centering
\includegraphics[width=1.1\columnwidth]{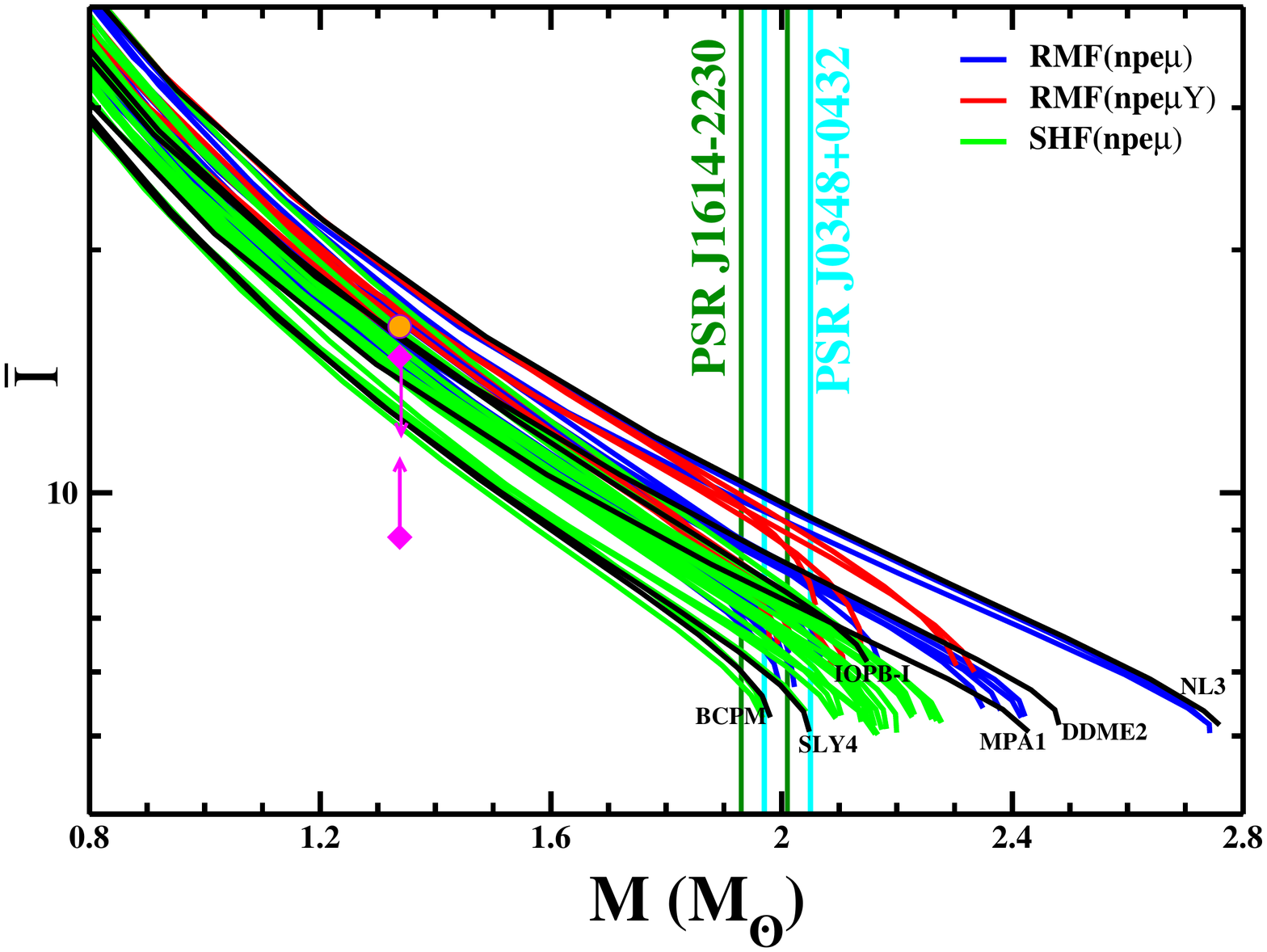}
\caption{The dimensionless moment of inertia $\bar{I}$ as a function of neutron star mass $M$ for various equations of state. The constraints on the moment of inertia of PSR J0737-3039A set by \citet{LVCeos}'s analysis of GW170817 are indicated by the overlaid arrows. The less restrictive upper bound from the original analysis of \citet{LVCdetection} is shown with a circle. The masses of the heaviest known neutron stars \citep{Demorest10,Antoniadis13} are included for reference. We have highlighted a representative subset of the equations of state, namely (in rough order of increasing stiffness) SLY4, BCPM, MPA1, DDME2, IOPB-I and NL3.}
\label{fig:mi}
\end{figure}

Our moment of inertia inference is the first use of the combined I-Love and binary Love relations to translate observations of a neutron star from one system to constraints on the properties of a neutron star from another. The constraints on $I_{\star}$ are displayed in Fig.~\ref{fig:mi} alongside the moments of inertia predicted by various candidate equations of state. We observe that the gravitational-wave data favor small moments of inertia at $M=1.338\,M_{\odot}$, a feature of soft models. This is consistent with the inferences drawn from the tidal deformability itself \citep{LVCdetection,De,LVCeos}, as the $I_{\star}$ constraints are derived from the same underlying observational data.

\section{\label{sec:eos}Candidate equations of state.}

The binary Love and I-Love relations are calculated using a large set of candidate neutron-star equations of state based on relativistic mean-field (RMF) and Skyrme-Hartree-Fock (SHF) treatments of the nuclear microphysics. These models' coupling constants are fixed by fitting to experimental data on the structure of select finite nuclei and the saturation properties of bulk nuclear matter (see e.g.~\citep{Kumar_param,Kumar_ns}). They closely reproduce observed features of nuclear matter at both microscopic and macroscopic scales, including the neutron skin thickness \citep{Kumar_ns}, the specific energy of sub-saturation neutron matter \citep{Kumar_param}, and the masses and radii of astrophysical neutron stars \citep{Fortin}. Moreover, the RMF and SHF equations of state are causal and thermodynamically stable by construction \citep{Tuhin}. The 53 recently developed RMF and SHF models considered here each support a $1.93\,M_{\odot}$ star, a conservative lower bound on the maximum neutron star mass \citep{Demorest10,Antoniadis13}.

Our representative set of equations of state comprises the RMF $npe\mu$-matter models BKA20 \citep{Agrawal10}, BSP \citep{Bijay12}, IOPB-I \citep{Kumar_ns}, Model1 \citep{Mondal15}, MPA1 \citep{MPA1}, MS1b \citep{Muller96}, SINPA \citep{Mondal16}, BSR2, BSR6, FSUGarnet, FSUGold2, G3, GM1, NL3, NL$3{\omega \rho}$, TM1 with standard nonlinear interactions and higher-order couplings, and DD2, DDH$\delta$, DDME2 with density-dependent linear interactions; the hyperonic $npe\mu Y$-matter variants BSR2Y, BSR6Y, GM1Y, NL3Y, NL3Yss, NL$3{\omega \rho}$Y, NL$3{\omega \rho}$Yss, DD2Y, DDME2Y; and the SHF $npe\mu$-matter models BSk20, BSk21, BSk22, BSk23, BSk24, BSk25, BSk26, KDE0v1, Rs, SK255, SK272, SKa, SKb, SkI2, SkI3, SkI4, SkI5, SkI6, SkMP, SKOp, SLY230a, SLY2, SLY4, SLY9.\footnote{For details on the unattributed equations of state, we refer the reader to \citet{Fortin,Kumar_param,Tuhin} and references therein.} The BCPM model \citep{Sharma15},
based on modern microscopic calculations using the Argonne $v_{18}$ potential plus three-body forces computed with the Urbana model, is also considered.

Many of these equations of state are \emph{unified}---they represent a single pressure-density relation that applies from the crust of the neutron star to its core. For these models, the outer crust is described by the BPS model \citep{Baym71}, and the inner crust equation of state is obtained with either a Thomas-Fermi calculation \citep{Grill14} (RMF) or the compressible liquid drop model plus variational methods \citep{Fortin} 
(SHF). Some of the models (BSP, FSUGarnet, G3, IOPB-I, Model1, MPA1, MS1b, SINPA) are available as core equations of state only, in which case we affix an SLY4 crust at low densities.

For the calculation of the relevant neutron star properties, namely $I$ and $\Lambda$, we adopt a piecewise polytrope representation of the equation of state. Phenomenological parameterizations of this kind have been shown to accurately reproduce the properties of a wide range of candidate equations of state \citep{Read}. In a piecewise polytrope, the equation of state in the $i^{\text{th}}$ segment is $p(\rho) = K_i \rho^{\Gamma_i}$, where $p$ is the pressure, $\rho$ is the mass density, $\Gamma_i$ is the adiabatic index and $K_i$ is a constant of proportionality with dimensions of $[\text{density}]^{1-\Gamma_i}/c^2$. The total energy density in the $i^{\text{th}}$ segment is \mbox{$\mu(\rho) = \rho c^2 + p/(\Gamma_i-1)$.} Here we implement the specific parameterization of \citet{Read}, which joins a three-segment piecewise polytrope to a low-density crust. It fixes the dividing densities \mbox{$\rho_1 = 10^{14.7}~\text{g cm}^{-3}$,} \mbox{$\rho_2 = 10^{15.0}~\text{g cm}^{-3}$} between core segments, and has four free parameters: $p_1 = p(\rho_1)$, the pressure at the first dividing density; and $\Gamma_1$, $\Gamma_2$ and $\Gamma_3$, the adiabatic indices for each of the polytropic segments. We fit the piecewise polytrope model to the tabulated equation of state data using the procedure described in \citet{Read}. Details of these computations, and the resulting parameterizations, will appear elsewhere.\footnote{B. Kumar and P. Landry (in preparation).}

\section{\label{sec:struc}Neutron star properties.}

The moment of inertia and tidal deformability of a neutron star are calculated by numerically integrating a system of equations of stellar structure. The Tolman-Oppenheimer-Volkoff equations \citep{Tolman,Tolman39}

\begin{align} \label{tov}
\frac{dp}{dr} = - \frac{G(\mu+p)(m+4\pi r^3 p/c^2)}{c^2 r^2 f} , \qquad \frac{dm}{dr} = 4\pi r^2 \mu/c^2 ,
\end{align}
where $f(r) := 1-2Gm/c^2r$, determine the profiles of total energy density $\mu(r)$, pressure $p(r)$ and mass $m(r)$ throughout the star. The stellar radius $R$ is defined by the condition $p(R)=0$, and the star's mass is \mbox{$M:=m(R)$.} The moment of inertia is computed by solving Hartle's slow rotation equation \citep{Hartle}

\begin{align} \label{slowrot}
0 =& \, r f \frac{d^2 \omega}{dr^2} + 4\left[ f - \pi r^2 (\mu/c^2 + p/c^2) \right] \frac{d\omega}{dr} \nonumber \\ &- 16\pi r (\mu/c^2+p/c^2) \omega
\end{align}
for the frame-dragging function $\omega(r)$; its surficial value $\omega_s := \omega(R)$ fixes the moment of inertia via \mbox{$I = (1-\omega_s)c^2 R^3/2G $.} The tidal perturbation $\eta(r)$ of the star's spacetime metric is governed by the equation \citep{Landry_surf}

\begin{align} \label{tidperts}
r \frac{d\eta}{dr} + \eta(\eta - 1) + A \eta - B = 0 ,
\end{align}
where

\begin{subequations}
\begin{align}
A &= 2 f^{-1} \left[ 1-\frac{3Gm}{c^2r} -2\pi r^2(\mu/c^2 + 3p/c^2) \right] , \\
B &= f^{-1} \left[ 6 - 4\pi r^2 (\mu/c^2+p/c^2)\left(3+\frac{d\mu}{dp}\right) \right] .
\end{align}
\end{subequations}
The tidal deformability is related to the surficial value $\eta_s := \eta(R)$ of the pertubation through

\begin{align}  \label{tidlns}
\Lambda =& \frac{\eta_s - 2 - 4C/f_s}{3C^{5}\left[R \; dF/dr - \left(\eta_s + 3 -4C/f_s \right)F \right]} ,
\end{align}
where $C := GM/c^2R$, $f_s := f(R)$, and \linebreak \mbox{$F(r) := {_2}F_1(3,5,6,2GM/c^2r)$} is a hypergeometric function; $F$ and $dF/dr$ are evaluated at $r=R$.

With a specification of the equation of state to close the system, Eqs.~\eqref{tov}-\eqref{tidperts} are integrated simultaneously from the center of the star, where the central density $\rho_c:= \rho(0)$ must be prescribed, to its surface. A sequence of stable neutron stars is constructed by sampling $\rho_c$ values up to $\rho_{\text{max}}$, the central density for which the stellar mass reaches a maximum $M_{\text{max}}$. Beyond $\rho_{\text{max}}$, the stars become unstable to radial perturbations \citep{Harrison}.

\begin{figure}
\includegraphics[width=1.1\columnwidth]{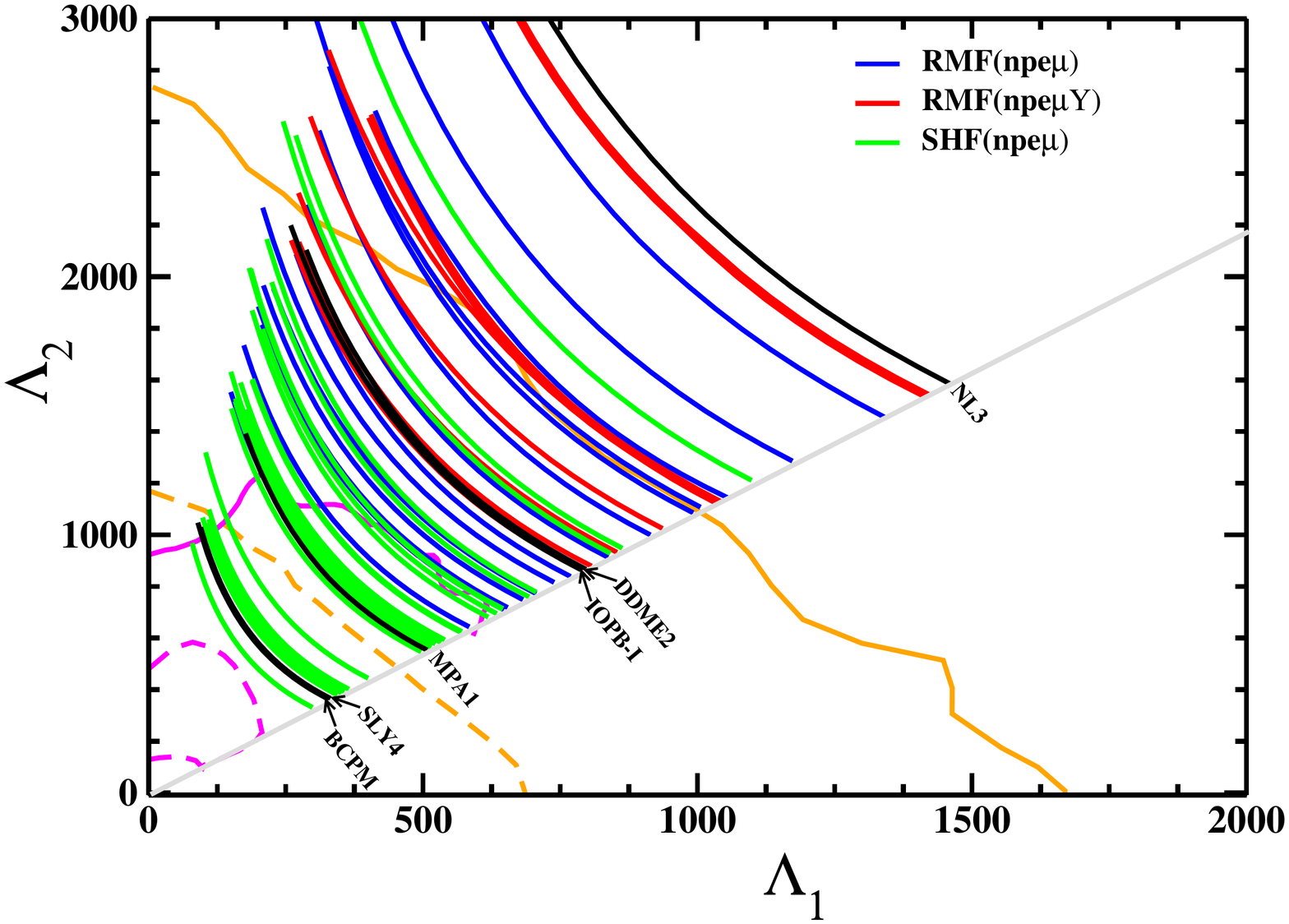} 
\caption{Tidal deformabilities for a $1.188\,M_{\odot}$ chirp-mass binary neutron star merger consistent with GW170817. The tidal deformabilities predicted by various candidate equations of state are evaluated for mass ratios $q \in [0.7,1.0]$, as inferred from the low-spin priors analysis of \citet{LVCdetection}. The 90\%-credible (50\%-credible) contours of the likelihood distribution determined by \citet{LVCeos}'s parameter estimation are shown pink and solid (dashed). The 90\%-credible (50\%-credible) contours from the original parameter estimation of \citet{LVCdetection}, which did not assume a universal equation of state, are shown orange and solid (dashed).}
\label{fig:bilove}
\end{figure}

Selecting 50 logarithmically spaced central densities in the interval $[1.0,8.5] \, \rho_{\text{nuc}}$ for the integrations, we compute a stable mass-sequence for each of our 53 equations of state. The moment of inertia and tidal deformability data obtained in this way are displayed in Figs.~\ref{fig:mi} and \ref{fig:bilove}. The dimensionless moment of inertia $\bar{I}(M)$ plotted in Fig.~\ref{fig:mi} is a decreasing function of the mass, with stiffer equations of state producing larger $\bar{I}$ at fixed $M$. Because Eq.~\eqref{slowrot} neglects rotational and tidal deformations of the neutron star, the moment of inertia we compute is that of a spherical star. The corrections to $I$ enter at second order in the dimensionless spin $\chi := c S/GM^2$ and at first order in the tidal perturbation $\epsilon := \delta R/R$. Since neutrons stars in binaries are expected to rotate slowly,\footnote{Pulsar A has the second-shortest rotational period of any known binary neutron star \citep{Burgay03}, surpassed only by the recently discovered double neutron star \citep{Stovall}. Theoretical modelling of its moment of inertia \citep{Bejger05,Morrison04} suggests that its dimensionless spin is $\chi \lesssim 0.05$ \citep{Damour12,Hannam}. This is confirmed in the Discussion.} spin corrections to the stellar structure are of order $\chi^2 \sim 10^{-3}$. Similarly, tidal corrections to the neutron star's shape are of order $(R/r_{12})^3 \sim 10^{-15}$, where $r_{12}$ is the binary separation.\footnote{The tidal deformation scales like $\delta R \sim \mathcal{E}^{\text{q}}_0 R^4/GM$ \citep{Landry_surf}, where $\mathcal{E}^{\text{q}}_0 \sim G q M/{r_{12}}^3$ is the quadrupolar tidal field, so $\epsilon \sim (R/{r_{12}})^3$ when the mass ratio $q \approx 1$. Taking \mbox{$R \approx 15~\text{km}$} and using Kepler's third law to relate $r_{12}$ to the orbital period and total mass of the system reported in \citet{Burgay03}, one obtains the estimate provided.}   We therefore neglect both $O(\chi^2)$ and $O(\epsilon)$ corrections to $I$ in this work.

Fig.~\ref{fig:bilove} presents the tidal deformability data $\Lambda(M)$ in the context of GW170817, showing the $\Lambda_{1}$ (respectively $\Lambda_{2}$) values corresponding to the 1.36-1.60$\,M_{\odot}$ high-mass component (1.17-1.36$\,M_{\odot}$ low-mass component)  of GW170817 for each equation of state. We have fixed the chirp mass $\mathcal{M} := (M_1 M_2)^{3/5}/(M_1 + M_2)^{1/5}$ to $1.188\,M_\odot$, the most likely value from the parameter estimation of \citet{LVCdetection}, and have allowed the mass ratio $q := M_2/M_1$ to run from $0.7$ to $1.0$, the inferred range of $q$ assuming small neutron star spins. We also plot the 90\% (50\%) credible contours for the parameter estimation of $\Lambda$ from \citet{LVCdetection,LVCeos} with solid (dashed) lines. The curves lying outside the contours are disfavored by GW170817; we observe that softer models, which produce smaller values of $\Lambda$ for fixed $M$, better match the observational data.

\section{\label{sec:ilove}Universal relations.}

\begin{figure}
\centering
\includegraphics[width=1.1\columnwidth]{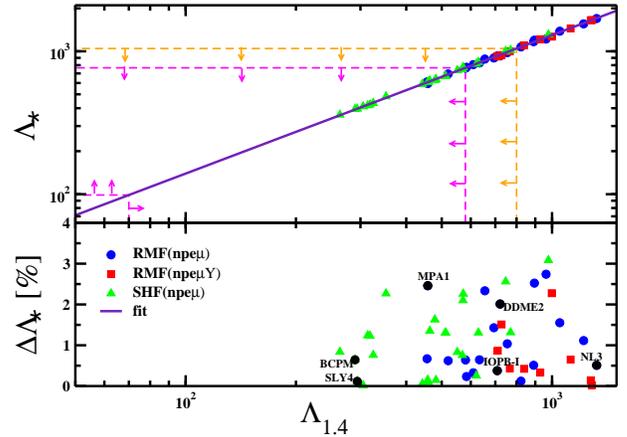}
\caption{Binary Love relation calculated with our set of 53 equations of state. The 90\%-credible gravitational-wave bounds on $\Lambda_{1.4}$ from \citet{LVCeos} (\citet{LVCdetection}), and the corresponding $\Lambda_{\star}$ constraints, are shown in pink (orange). Fit residuals are displayed in the lower panel.}
\label{fig:love}
\end{figure}

The binary Love and I-Love relations are calculated by performing log-log polynomial fits to the tidal deformability and moment of inertia data computed for the equations of state of interest. For the binary Love relation, the tidal deformabilities $\Lambda_{1.4} := \Lambda(1.4\,M_{\odot})$ and $\Lambda_{\star} := \Lambda(1.338\,M_{\odot})$ for each model are plotted against one another in Fig.~\ref{fig:love}. A fit to the relation

\begin{equation} \label{bilove}
\log_{10}{\Lambda_{\star}} = \sum_{n=0}^{1} a_n (\log_{10}{\Lambda_{1.4}})^n
\end{equation}
is performed, with the coefficients $a_n$ determined by least-squares regression (see Table~\ref{tb:coeff}). We have chosen a linear fit for the relation so as not to bias the extrapolation to the sparsely populated low-$\Lambda$ region of the plot, where the lower bound on the tidal deformability is located. Our fit is consistent with the $\Lambda_{\star}$-$\Lambda_{1.4}$ relation implied by the approximate scaling $\Lambda(M) \propto (R/M)^6$ identified in \citet{De}. As expected, the tidal deformabilities for all the equations of state hew closely to the fit, with deviations $\Delta \Lambda_{\star} = |\Lambda_{\star}-\Lambda_{\star}^{\text{fit}}|/\Lambda_{\star}^{\text{fit}} $ of no more than $3\%$. The fit residuals are plotted in the lower panel of Fig.~\ref{fig:love}.

\begin{table}
\centering
\caption{ Coefficients of the binary Love and I-Love fits, Eqs.~\eqref{bilove} and \eqref{ilove}.} \label{tb:coeff}
\begin{tabular}{cc}
\tablewidth{0pt}
\hline \hline
$\Lambda _{\star}\text{-}\Lambda _{1.4}$ & $\bar{I}\text{-$\Lambda $}$ \\ \hline
 $a_0 = 2.0592 \times 10^{-1}$ & $c_0 = 6.5022 \times 10^{-1}$ \\
 $a_1 = 9.6921 \times 10^{-1}$ & $c_1 = 5.8594 \times 10^{-2}$ \\
 - & $c_2 = 5.1749 \times 10^{-2}$ \\
 - & $c_3 = -3.6321 \times 10^{-3}$ \\
 -  & $c_4 = 8.5909 \times 10^{-5}$ \\
\hline
\end{tabular}
\end{table}

\begin{figure}
\includegraphics[width=1.1\columnwidth]{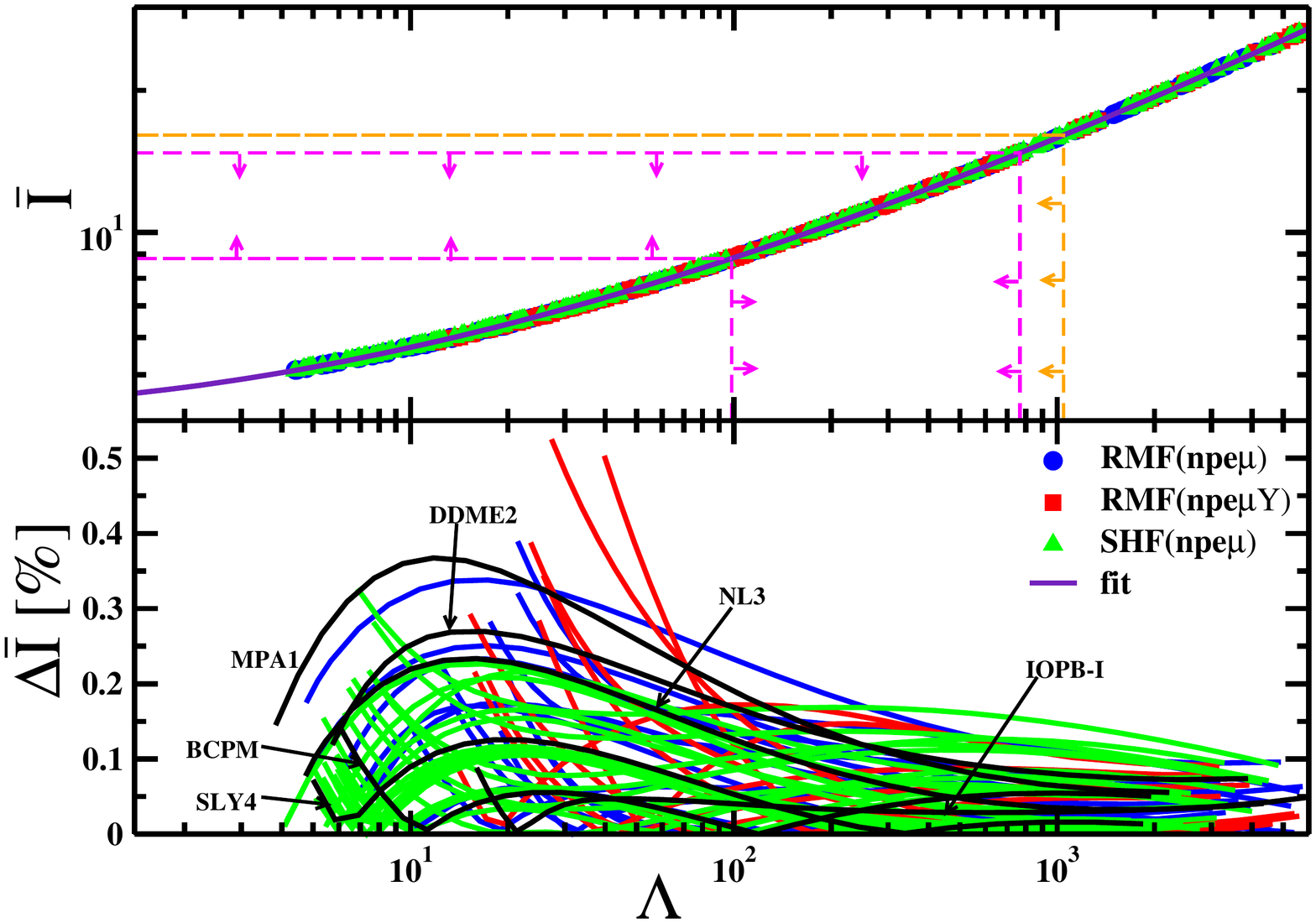}
\caption{I-Love relation calculated with our set of 53 equations of state. The $\Lambda_{\star}$ bounds inferred from GW170817 in Fig.~\ref{fig:love}, and the corresponding $\bar{I}_{\star}$ bounds,  are overlaid. Fit residuals are shown in the lower panel.}
\label{fig:ILQ}
\end{figure}

The I-Love relation is calculated in a similar fashion. For each equation of state in our set, the moment of inertia and tidal deformability mass-sequences (for \mbox{$M \in [M_{\odot}, M_{\text{max}}]$}) are plotted against one another, as shown in Fig.~\ref{fig:ILQ}, and a fit to the relation 

\begin{equation} \label{ilove}
\log_{10}{\bar{I}} = \sum_{n=0}^{4} c_n (\log_{10}{\Lambda})^n
\end{equation}
is performed, yielding the coefficients $c_n$ listed in Table~\ref{tb:coeff}. The deviations from the fit are also plotted in Fig.~\ref{fig:ILQ}, and do not exceed $0.6\%$ error. Our fit is nearly identical to the original I-Love relation calculated in \citet{YagiILQ}, though our residuals are slightly smaller because we omit the unrealistic polytropic models included there. We consider our recomputed fit to be more reliable than the original one because it is based on a larger, more representative set of equations of state.

Taken together, the binary Love and I-Love relations imply that

\begin{equation} \label{relation}
\log_{10}\bar{I}_{\star} = \sum_{n=0}^4 c_n (a_0 + a_1 \log_{10} \Lambda_{1.4})^n .
\end{equation}
Applying this formula to the $90\%$-credible $\Lambda_{1.4}$ constraints from GW170817, we obtain the bounds \linebreak \mbox{$\bar{I} = 11.10^{+3.64}_{-2.28}$}. (The upper bound derived from the initial analysis of GW170817 is $\bar{I} \leq 16.08$.) These correspond to the constraints on $I_{\star}$ given in the Introduction, after accounting for the uncertainty introduced by the dispersion in the universal relations. We model the error as a Gaussian centred on the fit with a symmetric two-sided $90\%$ confidence interval approximated by the maximum deviations from Figs.~\ref{fig:love} and \ref{fig:ILQ}. We have also verified that the results are insensitive to our choices of equations of state. If we use the 2\%-accurate I-Love relation of \citet{YagiILQ,YagiILQscience}, computed with a different set of models, in place of Eq.~\eqref{ilove}, the 90\% confidence interval on $I_{\star}$ is unchanged. The same is true if we repeat our inference using only the models that support e.g.~a $2.15\,M_{\odot}$ star.

\section{\label{sec:disc}Discussion.}

A precise astrophysical measurement of pulsar A's moment of inertia is expected within a few years. The measured value of $I_{\star}$ can be compared against the GW170817-based constraints presented here to test the universality of the neutron-star equation of state. If the neutron stars of GW170817 and PSR J0737-3039 hail from different populations---if, for instance, the former are SLY4-stars and the latter are quark stars---the model-independent mapping of Eq.~\eqref{relation} would break down, and the inference made here would be invalid.

It may also be possible to leverage the measured value of $I_{\star}$, if it can be determined with sufficient precision, to independently corroborate the waveform phase models used to extract the tidal deformability from the gravitational-wave signal. Analytic models of the tidal phase accumulate systematic error from omitted post-Newtonian point-particle and tidal terms \citep{Favata,Dietrich,Wade}. While one can control for the known omissions, the accuracy of current phase models would be confirmed if the gravitational-wave tidal constraints on $I_{\star}$ persistently agree with electromagnetic measurements as our knowledge of $\Lambda_{1.4}$ improves with further detections of neutron star mergers.

Taken in conjunction with pulsar A's known rotational frequency of 276.8 Hz, the moment of inertia bounds from GW170817 constrain the star's dimensionless spin to be $\chi = 0.020^{+0.007}_{-0.004}$ at 90\% confidence. (The upper bound on the spin from the initial analysis of GW170817 is $\chi \leq 0.029 $.) This suggests that pulsar A rotates slowly ($\chi \ll 1$), in keeping with expectations for binary neutron stars \citep{Damour12,Hannam}, \emph{provided that GW170817's neutron stars also rotated slowly}. This caveat is necessary because the $\Lambda_{1.4}$ constraints quoted in the Introduction were calculated under the assumption that $\chi \leq 0.05$ \citep{LVCdetection,LVCeos}. \citet{LVCdetection} presents an alternate upper bound of $\Lambda_{1.4} \leq 1400$ without the low-spin assumption. Repeating the inference with this revised bound, one finds $\chi \leq 0.034$. Hence, one can conclude on this basis that pulsar A spins slowly, given only the universality of the equation of state. We remark that this spin inference is essentially model-independent, since the universal relations are robust to the choice of EoSs used to compute them, and our error estimates for the fits---while weakly model-dependent---are practically negligible compared to the measurement uncertainty in $\Lambda$.

Finally, we point out that the moment of inertia inference performed here can be repeated for other systems. The recently discovered double neutron star PSR J1946+2052 has the shortest known orbital period for a binary of its kind \citep{Stovall}, making it another excellent target for a moment of inertia measurement. The method developed in this letter may be used to constrain its moment of inertia, and hence its spin, with $\sim 30 \%$ accuracy. Indeed, the method can be deployed across all systems for which a moment of inertia measurement is available as a systematic check of the equation of state's universality.

\acknowledgments

We gratefully acknowledge Jocelyn Read, John Friedman, James Lattimer, Reed Essick, S.~K.~Patra, Maya Fishbach and Zoheyr Doctor for useful comments about the manuscript.
This work was supported in part by the Natural Sciences and Engineering Research Council of Canada, and by NSF grants PHY 15-05124 and PHY 17-08081 to the University of Chicago.
B.~K. thanks the Navajbai Ratan Tata Trust, which also provided 
partial support for this work.

\bibliography{pwpolyeos-refs}

\begin{thebibliography}{}
\expandafter\ifx\csname natexlab\endcsname\relax\def\natexlab#1{#1}\fi
\providecommand{\url}[1]{\href{#1}{#1}}
\providecommand{\dodoi}[1]{doi:~\href{http://doi.org/#1}{\nolinkurl{#1}}}
\providecommand{\doeprint}[1]{\href{http://ascl.net/#1}{\nolinkurl{http://ascl.net/#1}}}
\providecommand{\doarXiv}[1]{\href{https://arxiv.org/abs/#1}{\nolinkurl{https://arxiv.org/abs/#1}}}

\bibitem[{{Aasi} {et~al.}(2015)}]{LIGO}
{Aasi}, J., {et~al.} 2015, CQGra, 32, 074001,
  \dodoi{10.1088/0264-9381/32/7/074001}

\bibitem[{{Abbott} {et~al.}(2017)}]{LVCdetection}
{Abbott}, B.~P., {et~al.} 2017, PhRvL, 119, 161101,
  \dodoi{10.1103/PhysRevLett.119.161101}

\bibitem[{{Abbott} {et~al.}(2018)}]{LVCeos}
---. 2018, arXiv.
\newblock \doarXiv{1805.11581}

\bibitem[{{Acernese} {et~al.}(2015)}]{VIRGO}
{Acernese}, F., {et~al.} 2015, CQGra, 32, 024001,
  \dodoi{10.1088/0264-9381/32/2/024001}

\bibitem[{{Agrawal}(2010)}]{Agrawal10}
{Agrawal}, B.~K. 2010, \prc, 81, 034323, \dodoi{10.1103/PhysRevC.81.034323}

\bibitem[{{Agrawal} {et~al.}(2012){Agrawal}, {Sulaksono}, \&
  {Reinhard}}]{Bijay12}
{Agrawal}, B.~K., {Sulaksono}, A., \& {Reinhard}, P.-G. 2012, NuPhA, 882, 1,
  \dodoi{10.1016/j.nuclphysa.2012.03.004}

\bibitem[{{Antoniadis} {et~al.}(2013)}]{Antoniadis13}
{Antoniadis}, J., {et~al.} 2013, Sci, 340, 448, \dodoi{10.1126/science.1233232}

\bibitem[{{Baym} {et~al.}(1971){Baym}, {Pethick}, \& {Sutherland}}]{Baym71}
{Baym}, G., {Pethick}, C., \& {Sutherland}, P. 1971, \apj, 170, 299,
  \dodoi{10.1086/151216}

\bibitem[{{Bejger} {et~al.}(2005){Bejger}, {Bulik}, \& {Haensel}}]{Bejger05}
{Bejger}, M., {Bulik}, T., \& {Haensel}, P. 2005, MNRAS, 364, 635,
  \dodoi{10.1111/j.1365-2966.2005.09575.x}

\bibitem[{{Burgay} {et~al.}(2003)}]{Burgay03}
{Burgay}, M., {et~al.} 2003, Natur, 426, 531, \dodoi{10.1038/nature02124}

\bibitem[{{Cahillane} {et~al.}(2017)}]{LIGOcalibration}
{Cahillane}, C., {et~al.} 2017, \prd, 96, 102001,
  \dodoi{10.1103/PhysRevD.96.102001}

\bibitem[{{Carney} {et~al.}(2018){Carney}, {Wade}, \& {Irwin}}]{Carney}
{Carney}, M.~F., {Wade}, L.~E., \& {Irwin}, B.~S. 2018, \prd, 98, 063004,
  \dodoi{10.1103/PhysRevD.98.063004}

\bibitem[{{Chatziioannou} {et~al.}(2018){Chatziioannou}, {Haster}, \&
  {Zimmerman}}]{Chatziioannou}
{Chatziioannou}, K., {Haster}, C.-J., \& {Zimmerman}, A. 2018, \prd, 97,
  104036, \dodoi{10.1103/PhysRevD.97.104036}

\bibitem[{{Damour} {et~al.}(2012){Damour}, {Nagar}, \& {Villain}}]{Damour12}
{Damour}, T., {Nagar}, A., \& {Villain}, L. 2012, \prd, 85, 123007,
  \dodoi{10.1103/PhysRevD.85.123007}

\bibitem[{{Damour} \& {Schafer}(1988)}]{Damour}
{Damour}, T., \& {Schafer}, G. 1988, NCimB, 101B, 127,
  \dodoi{10.1007/BF02828697}

\bibitem[{{De} {et~al.}(2018)}]{De}
{De}, S., {et~al.} 2018, PhRvL, 121, 091102,
  \dodoi{10.1103/PhysRevLett.121.091102}

\bibitem[{{Del Pozzo} {et~al.}(2013)}]{DelPozzo}
{Del Pozzo}, W., {et~al.} 2013, PhRvL, 111, 071101,
  \dodoi{10.1103/PhysRevLett.111.071101}

\bibitem[{{Demorest} {et~al.}(2010)}]{Demorest10}
{Demorest}, P.~B., {et~al.} 2010, Natur, 467, 1081, \dodoi{10.1038/nature09466}

\bibitem[{{Dietrich} {et~al.}(2018)}]{Dietrich}
{Dietrich}, T., {et~al.} 2018, arXiv.
\newblock \doarXiv{1804.02235}

\bibitem[{{Favata}(2014)}]{Favata}
{Favata}, M. 2014, PhRvL, 112, 101101, \dodoi{10.1103/PhysRevLett.112.101101}

\bibitem[{{Flanagan} \& {Hinderer}(2008)}]{Flanagan}
{Flanagan}, {\'E}.~{\'E}., \& {Hinderer}, T. 2008, \prd, 77, 021502,
  \dodoi{10.1103/PhysRevD.77.021502}

\bibitem[{{Fortin} {et~al.}(2016)}]{Fortin}
{Fortin}, M., {et~al.} 2016, \prc, 94, 035804,
  \dodoi{10.1103/PhysRevC.94.035804}

\bibitem[{{Gorda}(2016)}]{Gorda}
{Gorda}, T. 2016, \apj, 832, 28, \dodoi{10.3847/0004-637X/832/1/28}

\bibitem[{{Grill} {et~al.}(2014)}]{Grill14}
{Grill}, F., {et~al.} 2014, \prc, 90, 045803,
  \dodoi{10.1103/PhysRevC.90.045803}

\bibitem[{{Hannam} {et~al.}(2013)}]{Hannam}
{Hannam}, M., {et~al.} 2013, \apj, 766, L14,
  \dodoi{10.1088/2041-8205/766/1/L14}

\bibitem[{{Harrison} {et~al.}(1965){Harrison}, {Thorne}, {Wakano}, \&
  {Wheeler}}]{Harrison}
{Harrison}, B.~K., {Thorne}, K.~S., {Wakano}, M., \& {Wheeler}, J.~A. 1965,
  {Gravitation Theory and Gravitational Collapse} (Chicago, USA: University of
  Chicago Press)

\bibitem[{{Hartle}(1967)}]{Hartle}
{Hartle}, J.~B. 1967, \apj, 150, 1005, \dodoi{10.1086/149400}

\bibitem[{{Hinderer} {et~al.}(2010){Hinderer}, {Lackey}, {Lang}, \&
  {Read}}]{Hinderer}
{Hinderer}, T., {Lackey}, B.~D., {Lang}, R.~N., \& {Read}, J.~S. 2010, \prd,
  81, 123016, \dodoi{10.1103/PhysRevD.81.123016}

\bibitem[{{Kramer} \& {Wex}(2009)}]{KramerWex}
{Kramer}, M., \& {Wex}, N. 2009, CQGra, 26, 073001,
  \dodoi{10.1088/0264-9381/26/7/073001}

\bibitem[{{Kramer} {et~al.}(2006)}]{Kramer}
{Kramer}, M., {et~al.} 2006, Sci, 314, 97, \dodoi{10.1126/science.1132305}

\bibitem[{{Kumar} {et~al.}(2018){Kumar}, {Patra}, \& {Agrawal}}]{Kumar_ns}
{Kumar}, B., {Patra}, S.~K., \& {Agrawal}, B.~K. 2018, \prc, 97, 045806,
  \dodoi{10.1103/PhysRevC.97.045806}

\bibitem[{{Kumar} {et~al.}(2017){Kumar}, {Singh}, {Agrawal}, \&
  {Patra}}]{Kumar_param}
{Kumar}, B., {Singh}, S.~K., {Agrawal}, B.~K., \& {Patra}, S.~K. 2017, NuPhA,
  966, 197, \dodoi{10.1016/j.nuclphysa.2017.07.001}

\bibitem[{{Landry} \& {Poisson}(2014)}]{Landry_surf}
{Landry}, P., \& {Poisson}, E. 2014, \prd, 89, 124011,
  \dodoi{10.1103/PhysRevD.89.124011}

\bibitem[{{Lattimer} \& {Schutz}(2005)}]{Lattimer05}
{Lattimer}, J.~M., \& {Schutz}, B.~F. 2005, \apj, 629, 979,
  \dodoi{10.1086/431543}

\bibitem[{{Lyne} {et~al.}(2004)}]{Lyne04}
{Lyne}, A.~G., {et~al.} 2004, Sci, 303, 1153, \dodoi{10.1126/science.1094645}

\bibitem[{{Malik} {et~al.}(2018)}]{Tuhin}
{Malik}, T., {et~al.} 2018, PhRvC, 98, 035804,
  \dodoi{10.1103/PhysRevC.98.035804}

\bibitem[{{Messick} {et~al.}(2017)}]{LIGOsearch1}
{Messick}, C., {et~al.} 2017, \prd, 95, 042001,
  \dodoi{10.1103/PhysRevD.95.042001}

\bibitem[{{Mondal} {et~al.}(2015){Mondal}, {Agrawal}, \& {De}}]{Mondal15}
{Mondal}, C., {Agrawal}, B.~K., \& {De}, J.~N. 2015, \prc, 92, 024302,
  \dodoi{10.1103/PhysRevC.92.024302}

\bibitem[{Mondal {et~al.}(2016)Mondal, Agrawal, De, \& Samaddar}]{Mondal16}
Mondal, C., Agrawal, B.~K., De, J.~N., \& Samaddar, S.~K. 2016, \prc, 93,
  044328, \dodoi{10.1103/PhysRevC.93.044328}

\bibitem[{{Morrison} {et~al.}(2004){Morrison}, {Baumgarte}, {Shapiro}, \&
  {Pandharipande}}]{Morrison04}
{Morrison}, I.~A., {Baumgarte}, T.~W., {Shapiro}, S.~L., \& {Pandharipande},
  V.~R. 2004, ApJL, 617, L135, \dodoi{10.1086/427235}

\bibitem[{{M{\"u}ller} \& {Serot}(1996)}]{Muller96}
{M{\"u}ller}, H., \& {Serot}, B.~D. 1996, NuPhA, 606, 508,
  \dodoi{10.1016/0375-9474(96)00187-X}

\bibitem[{M{\"u}ther {et~al.}(1987)M{\"u}ther, Prakash, \& Ainsworth}]{MPA1}
M{\"u}ther, H., Prakash, M., \& Ainsworth, T. 1987, PhLB, 199, 469 ,
  \dodoi{https://doi.org/10.1016/0370-2693(87)91611-X}

\bibitem[{{Nitz} {et~al.}(2017)}]{LIGOsearch2}
{Nitz}, A.~H., {et~al.} 2017, \apj, 849, 118, \dodoi{10.3847/1538-4357/aa8f50}

\bibitem[{Oppenheimer \& Volkoff(1939)}]{Tolman39}
Oppenheimer, J.~R., \& Volkoff, G.~M. 1939, PhRv, 55, 374,
  \dodoi{10.1103/PhysRev.55.374}

\bibitem[{{{\"O}zel} \& {Freire}(2016)}]{OzelFreire}
{{\"O}zel}, F., \& {Freire}, P. 2016, ARA\&A, 54, 401,
  \dodoi{10.1146/annurev-astro-081915-023322}

\bibitem[{{Pankow} {et~al.}(2018)}]{LIGOglitch2}
{Pankow}, C., {et~al.} 2018, arXiv.
\newblock \doarXiv{1808.03619}

\bibitem[{{Raithel} {et~al.}(2016){Raithel}, {{\"O}zel}, \&
  {Psaltis}}]{Raithel}
{Raithel}, C.~A., {{\"O}zel}, F., \& {Psaltis}, D. 2016, \prc, 93, 032801,
  \dodoi{10.1103/PhysRevC.93.032801}

\bibitem[{{Read} {et~al.}(2009){Read}, {Lackey}, {Owen}, \& {Friedman}}]{Read}
{Read}, J.~S., {Lackey}, B.~D., {Owen}, B.~J., \& {Friedman}, J.~L. 2009, \prd,
  79, 124032, \dodoi{10.1103/PhysRevD.79.124032}

\bibitem[{{Read} {et~al.}(2013)}]{ReadGW}
{Read}, J.~S., {et~al.} 2013, \prd, 88, 044042,
  \dodoi{10.1103/PhysRevD.88.044042}

\bibitem[{{Sharma} {et~al.}(2015)}]{Sharma15}
{Sharma}, B.~K., {et~al.} 2015, A\&A, 584, A103,
  \dodoi{10.1051/0004-6361/201526642}

\bibitem[{{Stovall} {et~al.}(2018)}]{Stovall}
{Stovall}, K., {et~al.} 2018, ApJL, 854, L22, \dodoi{10.3847/2041-8213/aaad06}

\bibitem[{{Tolman}(1939)}]{Tolman}
{Tolman}, R.~C. 1939, PhRv, 55, 364, \dodoi{10.1103/PhysRev.55.364}

\bibitem[{{Veitch} {et~al.}(2015)}]{LIGOpe}
{Veitch}, J., {et~al.} 2015, \prd, 91, 042003,
  \dodoi{10.1103/PhysRevD.91.042003}

\bibitem[{{Wade} {et~al.}(2014)}]{Wade}
{Wade}, L., {et~al.} 2014, \prd, 89, 103012, \dodoi{10.1103/PhysRevD.89.103012}

\bibitem[{{Wex}(1995)}]{Wex}
{Wex}, N. 1995, CQGra, 12, 983, \dodoi{10.1088/0264-9381/12/4/009}

\bibitem[{{Worley} {et~al.}(2008){Worley}, {Krastev}, \& {Li}}]{Worley}
{Worley}, A., {Krastev}, P.~G., \& {Li}, B.-A. 2008, \apj, 685, 390,
  \dodoi{10.1086/589823}

\bibitem[{{Yagi} \& {Yunes}(2013{\natexlab{a}})}]{YagiILQ}
{Yagi}, K., \& {Yunes}, N. 2013{\natexlab{a}}, \prd, 88, 023009,
  \dodoi{10.1103/PhysRevD.88.023009}

\bibitem[{{Yagi} \& {Yunes}(2013{\natexlab{b}})}]{YagiILQscience}
---. 2013{\natexlab{b}}, Sci, 341, 365, \dodoi{10.1126/science.1236462}

\bibitem[{{Yagi} \& {Yunes}(2016)}]{YagiBiLove}
---. 2016, CQGra, 33, 13LT01, \dodoi{10.1088/0264-9381/33/13/13LT01}

\bibitem[{{Yagi} \& {Yunes}(2017{\natexlab{a}})}]{YagiILQreview}
---. 2017{\natexlab{a}}, PhR, 681, 1, \dodoi{10.1016/j.physrep.2017.03.002}

\bibitem[{{Yagi} \& {Yunes}(2017{\natexlab{b}})}]{YagiBiLove2}
---. 2017{\natexlab{b}}, CQGra, 34, 015006,
  \dodoi{10.1088/1361-6382/34/1/015006}

\end{thebibliography}

\end{document}